\documentstyle[12pt,aasms4]{article}

\begin{document}

\title{The Reionization History in the Lognormal Model}

\author{Jiren Liu\altaffilmark{1}, Li-Zhi Fang\altaffilmark{2}, Long-Long
Feng\altaffilmark{1,3},and Hong-Guang Bi \altaffilmark{2}}

\altaffiltext{1}{Center for Astrophysics, University of Science
and Technology of China, Hefei, Anhui 230026, China}
\altaffiltext{2}{Department of Physics, University of Arizona,
Tucson, AZ 85721} \altaffiltext{3}{National Astronomical
Observatories, Chinese Academy of Science, Chao-Yang District,
Beijing 100012, China}

\begin{abstract}
We study the evolution of baryonic gas before the reionization in
the lognormal (LN) model of cosmic clustering. We show that the
thermal history of the universe around the reionization can
roughly be divided into three epochs: 1) cold dark age $z>z_r$, in
which baryon gas is neutral, and opaque to Ly$\alpha$ photons; 2)
hot dark age $z_r > z> z_{gp}$, in which a predominant part of
baryon gas is ionized and hot, but it is still opaque to
Ly$\alpha$ photons; 3) bright age $z<z_{gp}$, in which the
universe is ionized highly enough to be transparent to Ly$\alpha$
photons. In the flat cold dark matter cosmological models given by
WMAP and COBE, the difference of the two redshifts $z_r - z_{gp}$
is found to be as large as $\sim 10$ with $z_r\sim 17$ and
$z_{gp}\sim 7$. This reionization history naturally yields a high
optical depth to the CMB $\tau_e \simeq 0.12 - 0.19$ observed by
the TE polarization of the WMAP, and a low redshift $z_{gp}$ of
the appearance of the Ly$\alpha$ Gunn-Peterson trough $z_{gp}
\simeq 6 - 8$ in QSO's absorption spectra. The reason why the
universe stays long in an ionized, yet Ly$\alpha$ opaque, stage is
because the first photo-ionization heats the intergalactic gas
effectively and has balanced the gravitational clustering a long
period of time. Therefore, the result of a high $\tau_e$ and low
$z_{gp}$ is a common feature of all the models considered. Besides
the cosmological parameters, the only free parameter we used in
the calculation is $N_{ion}$, the mean ionization photons produced
by each baryon in collapsed objects. We take it to be 40 - 80 in
the calculation.

\end{abstract}


\section{Introduction}

The large scale structures of the universe evolved from initial
mass density perturbations seeded by the quantum fluctuations of
inflation. The power spectrum of the initial perturbation is of
power law, $P(k) \propto k^n$, with index $n \sim 1$, and so the
mass density perturbation is stronger on smaller scales.
Accordingly, objects on smaller scales generally formed at earlier
time. Small objects like primordial black holes could form at the
early radiation-dominated era. Objects on scales of stars could
form in the dark age, i.e. between the recombination and the
reionization. These stars play the leading role of the history of
the reionization.

The study of structure formation in the cosmic dark age is challenged
recently by two observations. 1. The complete Ly$\alpha$
Gunn-Peterson trough in the absorption spectra of high redshift
quasars shows that the reionization occurred probably at the
redshift $z_r\simeq 6 - 8$ (Fan et al. 2002, and references
therein). 2. The TE polarization of the Wilkinson Microwave
Anisotropy Probe (WMAP) yields a model-independent value of the
electron scattering optical depth to the cosmic microwave
background (CMB), $\tau_e= 0.17\pm 0.04$, which requires the
reionization redshift to be $z \simeq 11 - 30$ if the reionization
happen rapidly (Kogut et al. 2003). Therefore, a problem naturally
arisen is how to reconcile the two results, or why the
reionization redshift from the Ly$\alpha$ spectrum is
significantly lower than that given by the optical depth to the
CMB. It has been realized that the high $\tau_e$ and low $z_r$
probably indicates that the reionization is not abrupt, but lasts
a long period from $z=30-20$ to $6-8$ (Cen 2003; Haiman \& Holder
2003; Whyithe \& Loeb 2003; Holder et al. 2003; Hui \& Haiman
2003; Onken \& Miralda-Escud\'e 2003). The further questions we
might ask are how to describe the long period of the reionization
and which is the physical mechanism leading to such a long period
of the reionization? This paper is to address these problems using
a semi-analytical approach in the lognormal (LN) model.

The LN model assumes that the mass field of intergalactic medium
(IGM) is given by an exponential mapping from the linear field (Bi
1993; Bi \& Davidsen 1997). The probability distribution function
(PDF) of the IGM field is then lognormal. The first argument in
supporting the lognormal PDF is probably from the isothermal model
of the IGM in the gravitational potential as
\begin{equation}
\rho({\bf x}) \propto \exp \left[-\frac{\mu m_p}{\gamma kT}
\phi_{dm}({\bf x}) \right].
\end{equation}
where $\rho({\bf x})$ is the IGM mass density and $\phi_{dm}({\bf
x})$ the gravitational potential. $T$ and $\mu$ are, respectively,
the temperature and molecular weight of the gas, and $\gamma$ is
the polytropic index.  Eq. (1) is the well-known IGM model of
intracluster gas (Sarazin \& Bahcall 1977). Zeldovich et al.
(1990) pointed that when the gravity potential $\phi_{dm}({\bf
x})$ is a random field with a Gaussian PDF, the field $\rho({\bf
x})$ of eq.(1) is lognormal. It has been shown that the lognormal
solution is a reasonable approximation of the dynamical equations
of the IGM (e.g. Jones 1999; Matarrese \& Mohayaee 2002).
Therefore, the LN is dynamically legitimate to model the weakly
nonlinear properties (Bi \& Davidsen 1997), and highly
non-Gaussian features (Feng \& Fang 2000) of the IGM. Recently it
is also used to study the collapsing of hydrogen clouds of the IGM
(Bi et al. 2003, here after BFFJ).

As has been emphasized in BFFJ, the formation of the first
generation of collapsed hydrogen clouds is given by rare events,
which is sensitive to the tail of the mass field PDF on the high
perturbation side. It is noted that the LN PDF is of long tailed
and predicts a higher probability of rare events of high
perturbation than PDFs without a long tail. Upon this
consideration, the BFFJ showed that the reionization lasts a
period from redshift 15 to about 10. However, the BFFJ did not
considered the reaction of the evolved IGM on the reionization.
During the reionization, the IGM underwent a dramatic evolution
from neutral to ionized hydrogen, from cold to hot gas, and from
low to high entropy states. Consequently, the clustering behavior
of the IGM should also undergo a remarkable evolution in the epoch
around the reionization. On the other hand, the clustering of the
underlying dark matter is not affected by the reionization.
Therefore, we must analyze the impact of the IGM evolution upon
the reionization history.

The LN model directly provides the spatial distributions of the IGM.
It is effective to reveal the reaction of the evolution of the IGM
on the reionization. We will show that a long period of the
recombination is a natural outcome of the LN model
if considering the evolution of the IGM at the reionization. A
long period of the reionization history generally yield a
high optical depth $\tau_e$ to the CMB and a low reionization
redshift to the Ly$\alpha$ forests. We will calculate these
quantities in the cold dark matter cosmological models.

The paper is outlined as follow. In \S 2 we will describe the LN
model of the reionization. The history of hydrogen reionization
will be discussed in \S 3. Finally \S 4 gives the discussion and
conclusion.

\section{The Lognormal Model}

\subsection{Mass Field of Baryonic Gas}

In the LN model, the mass field of the IGM is given by an
exponential mapping of the linear density field of the underlying
dark matter as
\begin{equation}
\rho({\bf x}) = \bar{\rho}\exp[\delta_0({\bf x}) -  \sigma_0^2/2],
\end{equation}
where $\sigma_0$ is the variance of the linear Gaussian field of
the dark matter filtered on scales of the Jeans length $\lambda_b$
of the IGM; $\delta_0({\bf x})$ is the linear density contrast of
dark matter $\delta_{dm}({\bf x})$ smoothed by a window on the
Jeans scale, i.e. its Fourier counterpart is (Bi \& Davidsen 1997)
\begin{equation}
\delta_0({\bf k}) =\frac{\delta_{dm}({\bf k})}{1+x_b^2k^2}
\end{equation}
where $x_b$ is $1/2\pi$ of the comoving Jeans length of the IGM.

Obviously, when the fluctuation $\delta_0({\bf x})$ is small,
i.e. $\delta_0 ({\bf x}) \ll 1$, eq.(2)
reduces to approximately $\delta({\bf x})\simeq \delta_0({\bf x})$,
where $\delta({\bf x})=[\rho({\bf x})-\bar{\rho}]/\bar{\rho}$ is
the density contrast of the IGM. Moreover, when the
fluctuation $\delta_0({\bf x})$ is large, combining eqs.(2) and
(3) yields eq.(1) (Bi \& Davidsen 1997). Since the initial density
perturbation is stronger on smaller scales, the perturbations of
the dark matter gravitational potential $\phi_{dm}({\bf x})$ are
larger on smaller scales. Thus, even at early universe the IGM
field on small scales may already be in the state described by eq.(1). Thus,
it would be not appropriate to assume that the IGM field is always Gaussian
on all scales. The LN model provides a uniform description of the IGM
distribution in the linear, weakly nonlinear and lognormal regime
at various scales and redshifts.

Since the random field $\delta _0({\bf x})$ is Gaussian, the PDF
of the field $\rho({\bf x})$ is lognormal as
\begin{equation}
p(\rho/\bar{\rho})=\frac{1}{(\rho/\bar{\rho}) \sigma_0\sqrt{2\pi}}
 \exp\left [ -\frac{1}{2}
  \left(\frac{\ln (\rho/\bar{\rho}) + \sigma_0^2/2}{\sigma_0}\right )^2
   \right ], \hspace{3mm} \rho \geq 0.
\end{equation}
which gives the probability density of the event such that the mass
density within an area on scales of the Jeans length is found to
be equal to $\sim \rho/\bar{\rho}$. The tail of the PDF eq.(4) is
$\sim \exp[-(1/2)(\ln \rho/\bar{\rho})^2/\sigma_0^2+\ln
\rho/\bar{\rho}]$. Therefore, the probability of $\rho/\bar{\rho}$
for a given $\sigma_0$ would be equal to the probability of
$(\rho/\bar{\rho})^2$ for $2\sigma_0$. Namely, the change of
variance $\sigma_0$ by a factor 2 leads to a change of the
possible high density events from $\rho/\bar{\rho}$ to
$(\rho/\bar{\rho})^2$.

Obviously, the PDF is normalized as $\int_{0}^{\infty}p(x)dx =1$, and
the variance of $\rho/\bar{\rho}$ is given by (Vanmarcke 1983)
\begin{equation}
\sigma=[e^{\sigma_0^2}-1]^{1/2}.
\end{equation}
The $n^{th}$ moment of $\rho$ is
\begin{equation}
\overline {(\rho/\bar{\rho})^n}=
   \exp \left [(n^2-n)\frac{\sigma_0^2}{2}\right ].
\end{equation}
Equation (6) shows $[\overline {(\rho/\bar{\rho}^n}]^{1/n} >
[\overline {(\rho/\bar{\rho})^2}]^{1/2}$. That is, the ratio between the
high- and low-order moments is divergent in the limit of
$n \rightarrow \infty$. A field with such a divergent moment ration, by
definition, is intermittent (Zel'dovich, Ruzmaikin, \& Sokoloff 1990).
Actually, LN is typically a mathematical model of intermittent field.
This property is in good consistent with the detected intermittency of
the transmitted flux of QSO Ly$\alpha$ absorption (Jamkhedkar, Zhan \&
Fang 2000; Feng, Pando \& Fang 2003, Jamkhedkar et al. 2003).

\subsection{Jeans Length of Baryonic Matter}

Equation (2) shows that, in the LN model, the clustering features
of the baryonic gas is characterized by the variance $\sigma_0$ of
the linear density field on the scale of the Jeans length.
Primordial baryons, created at the time of nucleosynthesis,
recombines with electrons to become neutral gas at $z \sim 1000$.
Thereafter, the gas cools down adiabatically with the expansion of
the universe. In a homogeneous universe with the mean mass density
$\rho_m$, the Jean length of gaseous baryonic matter or the IGM is
defined by $\lambda_b\equiv v_s(\pi/G\rho_m)^{1/2}$, where $v_s$
is the sound speed of gas. The corresponding Jeans mass is
$m_J=(4\pi/3)\lambda_b^3\rho_m$. In a comoving scale of
$x_b=\lambda_b/2\pi$, we have (Bi \& Davidsen 1997)
\begin{equation}
x_b =\frac{1}{H_0}\left[\frac{2\gamma k_BT}
{3\mu m_p\Omega_m(1+z)}\right]^{1/2}.
\end{equation}
The polytropic index is assumed to be $\gamma=5/3$, and hydrogen
temperature thus follows $T\propto\bar{\rho}^{2/3}$, where
$\bar{\rho}$ is the mean mass density of baryonic matter. At
$z=15$, the hydrogen temperature is $\sim 4$ K (Medvigy  \& Loeb
2001), and the Jeans length $x_b\sim 1.2 $ kpc. The redshift
evolution of the comoving Jeans length $x_b$ is approximately
given by $(1+z)^{1/2}$.

During the reionization, the baryonic gas will be heated by the UV
ionizing background from few degrees of temperature to $\sim 1.3
\times 10^4K$; and correspondingly, there is about 2 magnitudes
increasing in $x_b$ or 6 magnitudes in the Jeans mass (e.g.
Ostriker \& Gnedin 1996). After the reionization, the IGM
temperature is maintained at about $\sim 10^4K$ by the UV
background photons, and therefore, $x_b$ will gradually increase
with the decrease of $z$ due to the factor $(1+z)$ in eq.(7).

For the variance $\sigma_0$ in eq.(2), we consider three fiducial
cosmological models. Two are inferred from the WMAP, and one from
the COBE. The cosmological parameters of these models are listed in
Table 1. We use the linear power spectrum of the dark matter given
by Eisenstein \& Hu (1999). The linear variance is approximately
growing with redshift as $\sigma_0(z) \propto 1/(1+z)$. For
cosmological models of $\Omega=1$ and $\Omega_{\Lambda}\neq 0$,
the linear growth of density perturbations is not exactly
following the factor $1/(1+z)$ (Lahav et al. 1991). We take account of
this effect by replacing the factor $1/(1+z)$ by
$1/g(\Omega_m,\Omega_\Lambda)(1+z)$, where the $g$ factor is
\begin{equation}
g(\Omega_m,\Omega_\Lambda)=\frac{2.5\Omega_m}
  {\Omega_m^{4/7}-\Omega_\Lambda+(1+\Omega_m/2)
   (1+\Omega_\Lambda/70)}.
\end{equation}
This correction is approximately reasonable at $z>1$ (Avelino \&
de Carvalho 1999, Carroll et at. 1992). For the models of  WMAP1,
WMAP2 and  COBE, the $g$ factor is also shown in Table 1.

Because the powers of the linear perturbations $k^3P(k)$ in all of
these models are smaller on larger scales, the two order magnitude
jump of $x_b$ at the reionization generally leads to a drop-off of
the variance $\sigma_0$. A typical curve of $\sigma_0$ for the
model WMAP2 is displayed in Fig. 1, in which the reionization is
assumed to happen instantly around redshift $z_r=17$. The dotted
line in Fig. 1 is for $\sigma_0$ in case of no reionization
occurring. The zigzag $z$-dependence around the reionization is a
common feature of the $\sigma_0$ curve. For instance, the Fig. 2
of the BFFJ shows also a zigzag at $z = 7$. In the BFFJ, it is
{\it assumed} that the reionization takes place at redshift $z_r =
7$. In this paper, we will not use {\it a priori} zigzag redshift
$z_r$, but determine it by a self-consistent calculation. It
should be pointed out that the zigzag $z$-dependence of $\sigma_0$
does not affect the evolution of dark matter, of which the
clustering is basically regardless the reionization.

\begin{table}[t]
\caption{Optical depth in three cosmological models}
\bigskip
\begin{tabular}{llllllll}
\tableline
model & $\Omega_m$ & $\Omega_\Lambda$ & $\Omega_b$ &
      $n_s$ & $\sigma_8$ & $h$ & $g$ \\
\tableline
 WMAP1 & $0.27$ & $0.73$ & $0.044$ & $0.97$ & $0.84$ & $0.71$
  & 0.758 \\
 WMAP2 & $0.29$ & $0.71$ & $0.047$ & $0.99$ & $0.9$ & $0.72$
  & 0.772 \\
 COBE  & $0.3$ & $0.7$ & 0.045 & 1 & 0.85 & 0.7 & 0.797 \\
\tableline
\end{tabular}
\end{table}

\subsection{Cumulative Mass Fraction and Clumping Factor}

For a field with the PDF of eq.(4), the cumulative mass fraction
$M(>\rho/\bar\rho)$, i.e. the fraction of mass in regions having
mass density larger than a given $\rho/\bar\rho$, is
\begin{equation}
M[>(\rho/\bar{\rho})] = \int ^{\infty}_{\rho/\bar{\rho}} x p(x) dx
= \frac{1}{2} {\rm erfc}
\left [\frac{\ln (\rho/\bar{\rho})}{\sqrt{2} \sigma_0}-
\frac{\sigma_0}{2\sqrt{2}}\right ].
\end{equation}
which implies that $M[>(\rho/\bar{\rho})]$ is a strongly varying
function of the variance $\sigma_0$, but weak for the variable
$\rho/\bar{\rho}$. Fig. 2 demonstrates a cumulative mass fraction
for $\rho/\bar{\rho}=2, \ 6$ and 10, in which $\sigma_0$ is taken
from Fig. 1, and using the dotted line for $z<z_r$, i.e. without
considering the drop-off of $\sigma_0$ at $z_r$. Thus, Fig. 2 can
be only used at redshift $z>z_r$.

The decrease of $M(>\rho/\bar{\rho})$ with $\rho/\bar{\rho}$ is
much slower than a Gaussian PDF, or PDFs with a Gauss-like tail.
For a Gaussian tail, we have $M(>10)/M(>6) \simeq
\exp[(-10^2+6^2)/2\sigma^2_0]$, or $M(>10)/M(>6) < \exp(-30)$,
when $\sigma_0 \leq 1$. However, Fig. 2 shows that even when $z$
is as high as 20, $M(>10)$ is less than $M(>6)$ only by a factor
less than 10. This is obviously due to the long tail behavior of
the lognormal PDF, which gives a larger probability of high
density events at high redshifts when the variance $\sigma_0$ is
still small.

The variable $\rho/\bar{\rho}$ in eq.(9) is the density within an
area on scales of the Jeans length, and therefore, the area with
$\rho/\bar{\rho}>1$ has mass larger than the Jeans mass. However,
these regions are not always in collapsing. This is because that
the IGM clouds have no time to collapse even when the cloud mass
is larger than the Jeans mass. BFFJ found that at redshift $z
\simeq 7$, only clouds with density $\rho/\bar{\rho}> 6$ are in
the turning phase. The threshold $\rho/\bar{\rho} \simeq  6$ is
much larger than the Jeans mass. In other words, the clustering on
the Jeans length scales is significantly suppressed. The
suppression of collapsing on the Jeans length scales has also been
noted in studying the so-called filtering scale (Gnedin \& Hui
1998).

The threshold for the clouds going to collapsing at redshift $z>7$
is a little lower than 6. Therefore, the mass fraction of
collapsed clouds can be estimated by the cumulative mass fraction
$M(>6)$. One can estimated the velocity dispersion of the
collapsed clouds by considering that in the LN model the collapsed
clouds in potential well is probably isothermal [eq.(1)]. In this
case, the velocity dispersion of the collapsed clouds is
approximately (e.g. Jing \& Fang 1994)
\begin{equation}
\sigma_v \simeq 13 H_0 x_b(1+z)^{1/2}
\Omega_m^{1/2}(\rho/\bar{\rho})^{1/2}
\end{equation}
Thus, for $h=0.7$, $\Omega_m=0.3$, the velocity dispersion of the
collapsed clouds of $\rho/\bar{\rho}>6$ is from $>$ 3.8 km s$^{-1}$
at $z=10$ to $>$ 7.7 km s$^{-1}$ at $z=20$. If using
$\rho/\bar{\rho}>10$, we have $\sigma_v>$ 5.0 km s$^{-1}$ at $z=10$
to $>$ 9.8 km s$^{-1}$ at $z=20$. Therefore, all these clouds have
mass larger than $10^4$ M$_{\odot}$, the $H_2$ cooling can take
place.

The so-called clumping factor is the second moment of the IGM
density field defined by
\begin{equation}\label{}
C(z) \equiv \overline {(\rho/\bar{\rho})^2} = e^{\sigma_0^2(z)}.
\end{equation}
Fig. 3 plots the clumping factor $C(z)$ in the three models with
$\sigma_0$ without considering the drop-off at $z_r$. It can only
be used for redshift $z>z_r$. The clumping factor is significantly
$z$-dependent. For all three models, $C(z)$ are large only at
redshift $z<10$, and rapidly decreases to $<3$ at redshifts $z>
15$. At a first glance, it seems to be a contradict between a
higher probability of clustering and a lower clumping factor.
Actually, it is nothing but a typical feature of intermitted
field, which consists of high density spikes and low density
regions between the spikes, and therefore, dense objects can be
effectively formed in the spikes, while the mean clumping factor
over the whole space is relatively low.

\section{History of the Reionization}

\subsection{Clustering of Baryonic Gas around Reionization}

 From Fig. 1, one can see that the $z$-dependence of $\sigma_0$
before and after the reionization $z_r$ are similar. Before $z_r$,
$\sigma_0$ is increased from 0.4 to 0.8, and after $z_r$,
$\sigma_0$ is again to increase from about 0.4 to higher values.
A fall of $\sigma_0$ is by a factor of $\sim$ 2 at $z_r$. Of course,
the change of $\sigma_0$ at the drop is dependent on $z_r$. In
this section, we only qualitatively describe the effect
of the $\sigma_0$ drop on the clustering of baryonic gas.
Quantitative calculation will be given in next two subsections.

As emphasized in \S 2.1, the cumulative mass fraction is a
strong function of $\sigma_0$. The drop of $\sigma_0$ by a factor
of two leads to a substantial change of the cumulative mass fraction.
One can then divide the IGM clustering into two phases, before and
after redshift $z_r$. Before the reionization, structure formation
was weak when the variance $\sim 0.4$, and became more active when
the variance reached $\sim 0.8$. Just after $\sigma_0$ dropping,
the structure formation tended to inactive again as the variance reduced
to $\sim 0.4$. Soon after the variance growing up to $\sim 0.8$,
the structure formation was built up once more. Therefore, there are two
phases, in which the rate of structure formations are high. The early
phase is just before $z_r$, and the late phase is at the second time of
$\sigma_0 \geq 0.8$.

The physical reason for the two-phase scenario of structure
formation is obvious. Baryonic gas falls into dark matter gravity
wells and the first generation of stars forms before $z_r$.
Because the UV ionization photons background is gradually
established from the first generation stars, baryonic gas is
heated to temperature $10^4K$, and the Jeans length becomes
bigger. The irregularities or clustering within the big Jeans
scale would be smeared out by thermal motion. Since the density
fluctuations on the big Jeans scales would still be in the linear
regime, the process of the structure formation is significantly
suppressed even halted. At the end of this smearing out, some
ionized hydrogen will move away from the bottom of the gravity
wells of dark matter halos. Eq.(1) also shows that when the
temperature $T$ increases by a factor of 10$^4$, many nonlinear
structures will disappear. The clustering will recover till the
variance $\sigma_0$ growing again. In the second phase, the
objects form from ionized IGM, and their masses range in the order
of $10^{9-10}$ $M_{\odot}$.

In the two-phase scenario, the structure formation will be slowed
down just after the reionization. Consequently, there is no enough
UV photons produced by stars to maintain a fully ionized IGM, and
a significant fraction of HI remains in the IGM. In this case, the
IGM is opaque for Ly$\alpha$ photons. The IGM becomes transparent
for Ly$\alpha$ photons till the redshift at which the rate of star
formation grows again to be able to produce enough UV photons for
maintaining a highly ionized IGM. Therefore, we can define two
redshifts: (1) $z_r$, the redshift of reionization, which is the
first time when the IGM was fully (or almost fully) ionized and
heated up to temperature $\sim 10^4$ K; (2) $z_{gp}$, the redshift of
the Gunn-Peterson transparency, which is the starting time for IGM
entering into the state such that the volume-averaged fraction of
HI is less than 10$^{-4}$, and the Gunn-Peterson absorption
ceased. In a word, the clustering of the IGM is suppressed during
the period from $z_r$ to $z_{gp}$, and thus leads to the
reionization epoch lasting a long period from $z_r$ to $z_{gp}$.

The two-phase scenario is based on the $z$-$\sigma_0$ relation
indicated in Fig. 1. Obviously, this $z$-dependence of $\sigma_0$
is approximate, as we assumed that the temperature of the IGM
suddenly increased by a factor of 10$^4$. More precisely, the
sharp decline of $\sigma_0$ at $z_r$ should be replaced by a
soften transition from the top branch to the bottom branch in the
$z-\sigma_0$ plane. Nevertheless, the mechanism for a long period
reionization outlined above still works, as it essentially depends
only on the decrease of $\sigma_0$ with the heating and ionizing
IGM.

\subsection{The Redshift of Reionization}

Let us first calculate the reionization redshifts $z_r$. The
reionization of hydrogen clouds is usually characterized by the
filling factor of ionized hydrogen $Q_{HII}(z)$. The reionization
redshift $z_r$ can approximately determined by requiring
$Q_{HII}(z_r) \geq 1$. The equation of $Q_{HII}(z)$ is given by
(e.g. Barkana \& Loeb 2001)
\begin{equation}
\frac{dQ_{HII}}{dt}=\frac{N_{ion}(t)}{0.76}
  \frac{dM(>\rho/\bar{\rho})}{dt}-
    \frac{1}{a^3}\alpha_B C_{HII}(t)\bar{n}_{H}^0 Q_{HII},
\end{equation}
where the factor $0.76$ is from the fraction of hydrogen in the
cosmic baryons, $\bar{n}_H^0 = 1.88 \times 10^{-7}
(\Omega_b$h$^2$/0.022) cm$^{-3}$ is the present value of the mean
number density of hydrogen, $N_{ion}(t)$ is the mean ionization
photons produced by each baryon in collapsed objects and might
be $t$-dependent. The first term on the r.h.s. of eq.(12) accounts
for the ionizing source. The ionization photons produced by
primordial black holes are negligible because of the spectral index
$n \leq 1$ on small scales (He \& Fang, 2002). The second term on
the r.h.s. of eq.(12) is from the recombination, in which $a$ is
the cosmic factor, $\alpha_B$ is the case B recombination
coefficient $\simeq 2.6 \times 10^{-13} cm^3s^{-1}$ for hydrogen
at $T = 10^4$ K, and $C_{HII}(t)$ is the clumping factor of $HII$.

Solving eq.(12) yields
\begin{equation}
Q_{HII}(t)=\int_0^t\frac{N_{ion}(t)}{0.76}
  \frac{dM(>\rho/\bar{\rho})}{dt'}e^{F(t',t)}dt',
\end{equation}
where
\begin{equation}
F(t',t)=-\alpha_B\bar{n}_H^0\int_{t'}^t\frac{C_{HII}(t'')}{a^3(t'')}dt''.
\end{equation}
In the solution (13), an uncertain parameter is $N_{ion}$,
which depends on the atomic processes in star formation. As usual,
one can express this number by $N_{ion}=N_\gamma \times f_{star}
\times f_{esc}$, where $f_{star}$ - the fraction of collapsed
baryon that formed star; $N_{\gamma}$ - the ionization photon
number produced per baryon in star; and $f_{esc}$ - the fraction
of photon that escaped from galaxy. Generally, the first
generation stars are massive (Abel et al. 2002, Bromm et al.
2002), and their lifetime are much less than the Hubble time at
high z that we are interested. Thus, it would be reasonable to
approximate the photons $N_{\gamma}$ being produced at the same
redshift as collapsed clouds.

The number $N_{\gamma}$ is quite different for stars with
different mass and metallicity. With the IMF of Scalo (1998), one
has $N_\gamma \approx 4000$ for stars with metallicity $\sim
0.05Z_\odot$. For the Salpeter's IMF, $N_{\gamma}$ is $\sim$ 6000
in average for stars with masses of 1 to 100 M$_\odot$. For stars
with zero metallicity, the number $N_{\gamma}$ could be as high as
10$^5$ (Bromm et al. 2001, Venkatesan et al. 2003). Observational
estimations of $f_{star}$ and $f_{esc}$ generally yield
$f_{star}f_{esc}\sim10^{-2}$. To sketch the reionization history
driven by the mechanism discussed in \S 3.1, we use a simplest
model of $N_{ion}$, i.e. let  $N_{ion}$ to be constant
($z$-independent), and takes the value of 40, 60, and 80
respectively, which covers the possible range of $N_{ion}$.

Figure 4 plots the filling factor $Q_{HII}$ vs. redshift $z$ in
the three cosmological models with $N_{ion}=40$, in which we take
$C_{HII}(z)=C(z)$. The curve of the model WMAP2 is similar to the
COBE, while the WMAP1 shows a lower filling factor. This is
because both the spectral index $n_s$ and the normalization of
power spectrum $\sigma_8$ of the WMAP1 are smaller than  other two
models, and accordingly, the WMAP1 gives smaller variance
$\sigma_0$ on the Jeans length of the IGM. From Fig. 4, the
redshift of reionization is in the range $18> z_r >16$.

We also estimate $z_r$ by taking $N_{ion}$=40, 60, and 80, and
$\rho/\bar{\rho}=6, 8, 10$. The results are tabulated in Table 2.
In all the considered parameter-range, the redshift of
reionization $z_r$ is $>10$. This result is consistent with the
BFFJ, which also found $z_r>10$ but using different method. The
BFFJ's conclusion is based on the evolution of the mass density
and velocity profiles of hydrogen clouds. Clearly, both approaches
arrives at the same result that the reionization could occur
much earlier than the redshift of an appearance of the
complete Gunn-Peterson trough of Ly$\alpha$ absorption.

\begin{table}[t]
\caption{Reionization redshift ($z_r$)}
\bigskip
\begin{tabular}{l|lllll}
\tableline
 $\rho/\bar{\rho}$  & 6 & 6 & 6 & 8 & 10 \\
\tableline
 $N_{ion}$   & 40 & 60 & 80 & 40 & 40  \\
\tableline
 WMAP1  &15.8 &17.0 &17.4 & 14.0 &12.9 \\
 WMAP2  &18.0 &19.5 &20.4 & 16.0 &14.9 \\
 COBE   &17.8 &19.0 &20.0 & 15.9 &14.6 \\
\tableline
\end{tabular}
\end{table}

The assumption of $C_{HII}(z)=C(z)$ is correct only if the density
distribution of $HII$ follows the distribution of total IGM. Since
$HII$ may favor to locate nearby collapsed clouds, we should use a
biased relation $C_{HII}(z)=bC(z)$. The bias parameter $b$ can be
estimated by
\begin{equation}
b=\left [\frac{\overline{\rho^4}}{(\overline{\rho^2)}^2} \right]^{1/2}
   = e^{2\sigma_0^2},
\end{equation}
where we used eq.(6). Since $0.4 < \sigma_0^2 <0.8$ (Fig. 1), we have
$2 < b < 5$. These values of $b$ is consistent with the estimation by
the property of collapsed clouds (Mo \& White 1996, Fang \& Jing 1998).
Because the velocity dispersion of collapsed clouds is small [Eq.(10)],
the parameter $b$ should be in the above-mentioned range. Figure 5
plots the filling factor $Q_{HII}$ vs. redshift $z$ for the model
$WMAP2$, $\rho/\bar{\rho}=6$ and $N_{ion}=40$, but taking the
biased clumping factor of $HII$ with parameter $b=$ 1.5, 2 and 4.
The results show that the uncertainty given by the bias factor is
about the same as the uncertainty given by cosmological parameters
(Fig. 4). Therefore this uncertainty will not affect the
conclusions shown below.

\subsection{The Redshift of the Gunn-Peterson Transparency}

Although $Q_{HII}$ is $\simeq 1$ at $z_r$, it does not mean that
$Q_{HII}\simeq 1$ for all $z<z_r$, because $\sigma_0$ fell at
$z =z_r$. For a self-consistent calculation of $Q_{HII}$ in the
epoch of $z<z_r$, we should use the lower branch of $\sigma_0$ as
shown in Fig. 1.

The filling factor $Q_{HII}$ at $z< z_r$ can also be calculated
using equation (12). Figure 6 plots the results of $Q_{HII}$ vs.
redshift in the model WMAP2 with the parameters $(\rho/\bar{\rho},
N_{ion})$ of (6, 40), (6, 80), and (8, 40). Other models yield
similar curves of $Q_{HII}$. A common feature is that the
reionization $Q_{HII}$ drops to a level less than 1 after $z_r$,
and then keeps $\simeq 0.5$ in the epoch of redshift $z_r
>z>z_{gp}$. It is  due to the decrease of the star
formation rate when $\sigma_0$ drops back to $\sim 0.4$. When
$z\rightarrow z_{gp}$, $\sigma_0$ becomes higher again [Fig. 1],
and therefore, $Q_{HII}$ quickly approaches to 1.

Each curves of Fig. 6 show that there are two solutions to the
equation $Q_{HII}(z) \simeq 1$, the first is at $z_r>15$, and the
second is at $z_{gp}\sim 6- 8$. This corresponds to the two phases
of the star formation. During the period between the redshift
$z_r$ and $z_{gp}$, the IGM is in multi-phase. The IGM consists of
neutral and ionized hydrogen, HI and HII. Both components are
comparable, and opaque for Ly$\alpha$ photons. However, the
temperature of such multiphase IGM can be of the order of 10$^4$ K
in average. Therefore, it is self-consistent to apply the bottom
branch of the $\sigma_0$ of Fig. 1.

When $z\rightarrow z_{gp}$, the filling factor of neutral
hydrogen becomes small quickly. The IGM is then transparent for
Ly$\alpha$ photons when the volume filling factor of
HI is less than 10$^{-4}$, and thus the redshift $z_{gp}$
corresponds to the onset of the Gunn-Peterson transparency.

It should be pointed out that the two peak curve of $Q_{HII}$
actually depends on the sharply dropping of $\sigma_0$ at $z_r$
(Fig. 1). As has been discussed in \S 3.1, more realistic model
should replace the sudden increase of IGM temperature by a soft
process. In this case, $Q_{HII}$ probably never reached 1 at
$z_r$, but at most shows a peak around $z_r$. Therefore, before
$z_{gp}$, the IGM has never been ionized to the level of
$1-Q_{HII} < 10^{-4}$. Thus, the redshift $z_{gp}$ is the first
time of the IGM becoming the Gunn-Peterson transparency. Table 3
gives $z_{gp}$ for three sets of parameters.

Figure 6 and Table 3 shows that the redshift interval $z_r -
z_{gp}$ is always as large as $\simeq 10 - 12$, which is not
sensitive to the adopted parameters. This is because long duration
from $z_r$ to $z_{gp}$  is essentially due to the zigzag
$z$-dependence of the $\sigma_0$, which is less-sensitive to
cosmological parameters, and the details of start formation. The
temperature of the IGM increases by a factor of about 10$^4$
during the reionization, while the entropy parameter
$T/n_{H}^{2/3}$ also increases by few orders. Since high
temperature and high entropy prevent baryonic gas from collapsing
into the gravity well of dark matter, the reionization is slowed
down with the increase of temperature and entropy of the IGM. It
behaves like a negative feedback mechanism: once the rate of star
formation is higher, the UV background produced by the star
formation is also higher, and then more baryonic gas will be
heated to higher temperature and higher entropy state. This
finally yields a suppression on the rate of star formation.
Therefore, the long duration of reionization is mainly given by
the evolution of the temperature and entropy of the baryonic gas
during the reionization.

\subsection{Optical Depth to the CMB}

If the reionization lasts the long period of $z_r$-$z_{gp}$, the
number density of free electrons is significant through the whole
epoch $z_r>z>z_{gp}$. One can then expect that in this
reionization history the electron scattering optical depth to the
CMB would be much larger than models with $z_r \simeq z_{gp}$.
With the solution of $Q_{HII}$, the electron scattering optical
depth to the CMB can be calculated by (e.g. Hui et al. 2003)
\begin{equation}
\tau_e=0.0525\times\Omega_bh\int_0^{z_{r}}
  \frac{V_{HII}(z)(1+z)^2dz}{\sqrt{\Omega_m(1+z)^3+\Omega_\Lambda}},
\end{equation}
where $V_{HII}(z)$ is the volume filling factor of $HII$. Eq.(16)
considered only the electrons from ionized hydrogen. To consider
the contribution of He, we had assumed the filling factor of HeII
to be identical to $HII$, i.e. the optical depth is increased by a
factor of 1.08. The HeII $\rightarrow$ $HeIII$ ionization occurred
at redshift $\simeq 3.3$ (Theuns et al. 2002), and therefore, the
correction of HeIII on $\tau_e$ is no more than 0.001.

The relation between the volume filling factor $V_{HII}$ and $Q_{HII}$
is dependent on the statistical behavior of the formation of the ionized
region.
If the ionized hydrogen region increase, $dQ_{HII}(t)$ in the duration $t$
to $t +dt$ is without spatial memory, the formation probability would be
uniform in space regardless whether there are ionized region formed at time
before $t$. Thus, we have equation $dV_{HII}/dt=(1-V_{HII})dQ_{HII}/dt$,
and the
solution is $V_{HII}(z)=1-\exp[-Q_{HII}(z)]$. However, the formation of
the ionized region probably is not without spatial memory. The probability of
forming collapsed clouds at the position where already hosted collapsed
clouds should be low. Therefore, the process of the ionized hydrogen region
formation most likely is self-avoid, i.e. the process has less probability
at the positions where the process has been visited(e.g. Cardy 1996). For a
perfectly self-avoiding process, no position can host the formation process
more than once (e.g. Cardy 1996). In this case, we have
$V_{HII}(z) =Q_{HII}(z)$ till $Q_{HII}\leq 1$.

Table 3 lists the results of $z_r$, $z_{gp}$ and $\tau_e$ for the
model WMAP2 with parameters $(N_{ion}, \rho/\bar{\rho})$ specified
by (6, 40), (8, 40) and (6, 80). The optical depth $\tau_e(1)$ is
for the model without any self-avoiding, while $\tau_e(2)$ is for
model with perfectly self-avoiding. Real number probably is in
between of these two ultimate models. All results shown in Table 3
give a low $z_{gp}$ $\sim 6 - 8$ and a high $\tau_e$ $\sim 0.12 -
0.19$. These results are excellently consistent with the
observations of the redshift of the Gunn-Peterson trough and the
electron optical depth to the CMB. The key of reconciling the two
observation is that there exists a long period of the universe to
have $0.1 < V_{HII}< 1-10^{-4}$. The free electrons in this period
contribute to a high value of $\tau_e$. On the other hand, the
Gunn-Peterson transparency occurred at the end of this epoch, and
the redshift $z_{gp}$ is small.

\begin{table}[t]
\caption{Reionization redshift, Gunn-Peterson redshift and optical
depth to CMB}
\bigskip
\begin{tabular}{ll|llll}
\tableline
 $\rho/\bar{\rho}$  & $N_{ion}$  & $z_{r}$ & $z_{gp}$ & $\tau_e(1)$
    & $\tau_e(2)$ \\
\tableline
 6 & 40  & 18.0 &7.0 & 0.150 & 0.181  \\
 8 & 40  & 16.0 &6.0 & 0.123 & 0.154 \\
 6 & 80 & 20.4 &7.9  & 0.159 & 0.189  \\
\tableline
\end{tabular}
\end{table}

\section{Discussion and Conclusions }

We showed with the LN model that the reionization and
Gunn-Peterson transparency emerged at very different time. The
thermal history of the universe around reionization can roughly be
divided into three epochs: 1.) cold dark age $z>z_r$, in which
baryonic gas is cold, and opaque for Ly$\alpha$ photons; 2.) hot
dark age $z_r > z> z_{gp}$, in which a substantial part of baryon
gas is ionized and hot, but it is still opaque for Ly$\alpha$
photons, 3.) bright age $z<z_{gp}$. For the LCDM models favored by
the WMAP, we have $z_r \simeq 13 - 19$ and $z_{gp}\simeq 6 -8$.
That is, the hot dark age is in the redshift interval from 13-19
to 6 -8. This thermal history naturally yields a high optical
depth $\tau_e=0.12 - 0.19$ and a low $z_{gp}=6 -8$, which is in
excellent agreement with the observations of $\tau_e$ (WMAP) and
$z_{gp}$ (Gunn-Peterson trough).

The long duration from $z_r$ to $z_{gp}$ is essentially due to
the evolution of the temperature and entropy of the baryonic gas
during the reionization. For all the considered cosmological
models, there is always a drop-off of $\sigma_0$ at $z_r$.
Therefore, a large difference $z_r- z_{gp}$ is a common property
of these models. The result of low $z_{gp}$ and high $\tau_e$
is less sensitive to the cosmological parameters.

This result is also weakly dependent on the details of production
of ionization photons from stars. This is due to the feedback
mechanism of the start formation (\S 3.3). The higher rate of
ionization photon production will cause a stronger input of entropy
to baryonic gas. It then leads to a lower rate of the collapsing of
baryonic clouds, suppressing the rate of star formation and
the UV photon production. Therefore, model, which properly
considers the dramatic evolution of the temperature and
entropy of the IGM, will generally yield a low $z_{gp}$ and a
high $\tau_e$.

The LN model provides an intuitive and quantitative description
of the effect of the IGM temperature evolution upon the reionization
history. This is because the LN model use directly variance $\sigma_0$
as a key parameter. It has already been recognized that the variance
$\sigma_0$ is more effective and successful in modeling the IGM
detected by QSO's Ly$\alpha$ forests (Bi \& Davidsen 1997). We showed
in this paper that the LN model provides a plausible description of
photoionized and neutral IGM not only in the redshift range $z\leq 4$,
but extended to $z\simeq 15$.

The hot dark age between the redshifts $z_r$ and $z_{gp}$ is
an interesting issue. During that period, star formation
is significant, although the rate of star formation is lower than that
at $z_r$. On the other hand, baryonic gas is in multiphase, and consists
of comparable components of HI and HII. The gas has been already hot.
Therefore, it is expected to find some observable signatures, such as
high redshift 21 cm emission, of the hot dark age.

\acknowledgments

JRL thanks Mr. Guopeng Shen and Jun Yu for useful discussions.
LLF acknowledges support from the National Science Foundation of
China (NSFC) and National Key Basic Research Science Foundation.

\newpage

\plotone{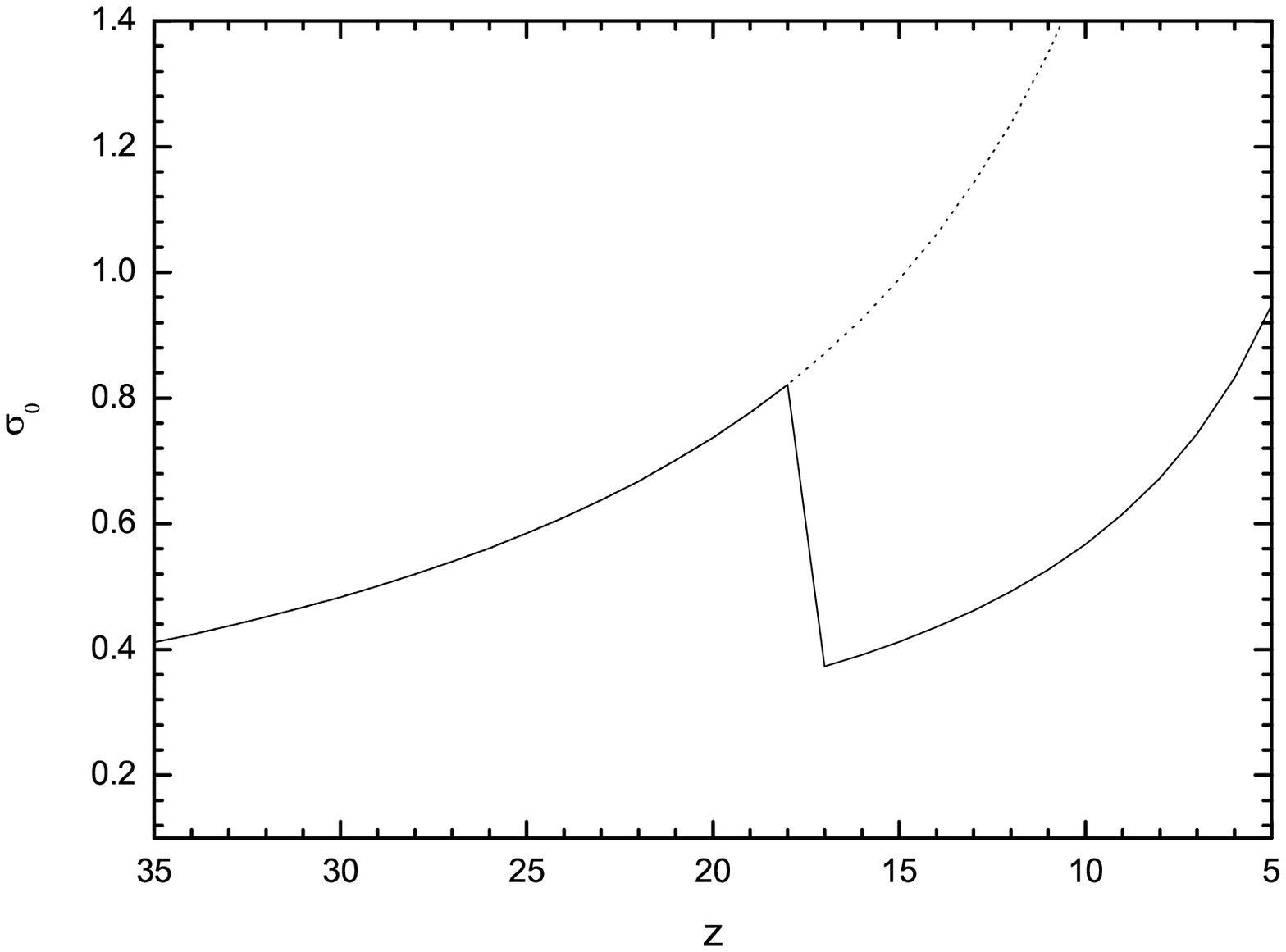} \figcaption {The variances of the linear density
fluctuations at the Jeans length as a function of redshift for
model WMAP2. The reionization redshift $z_r$ is taken to be 17.
The dot line at $z<z_r$ is for the variances without considering
the reionization. The solid line at $z<z_r$ is considered the
dropping of $\sigma_0$ at $z_r$ due to the increase of the
temperature of IGM.
 } \label{Fig1}

\plotone{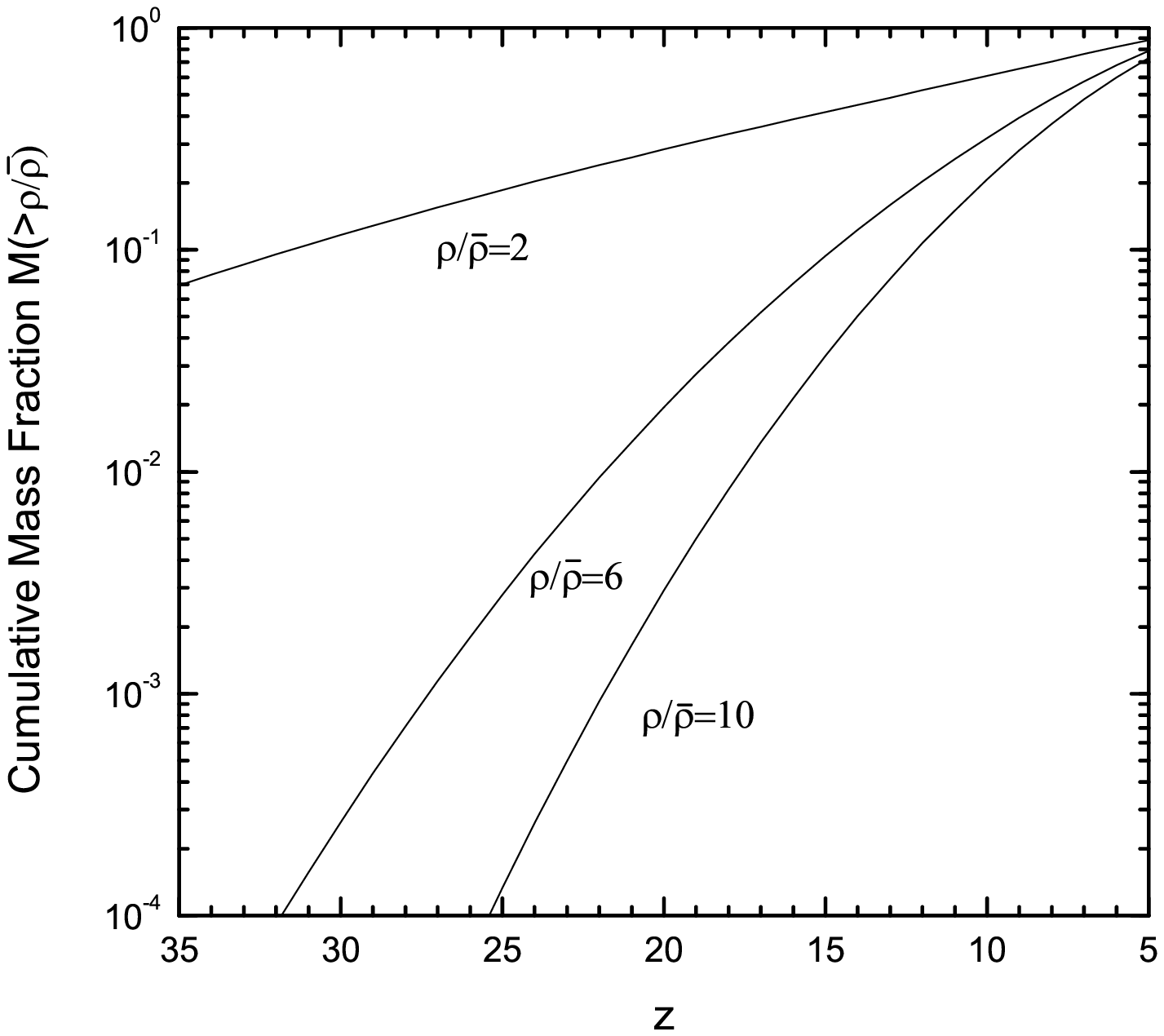} \figcaption {The cumulative mass fraction
$M(>\rho/\bar{\rho})$ for
 model WMAP2. $\rho/\bar{\rho}$ is taken to be 2, 6 and 10. At $z<z_r$,
the $\sigma_0$ is taken from the dot line in Fig. 1. } \label{Fig2}

\plotone{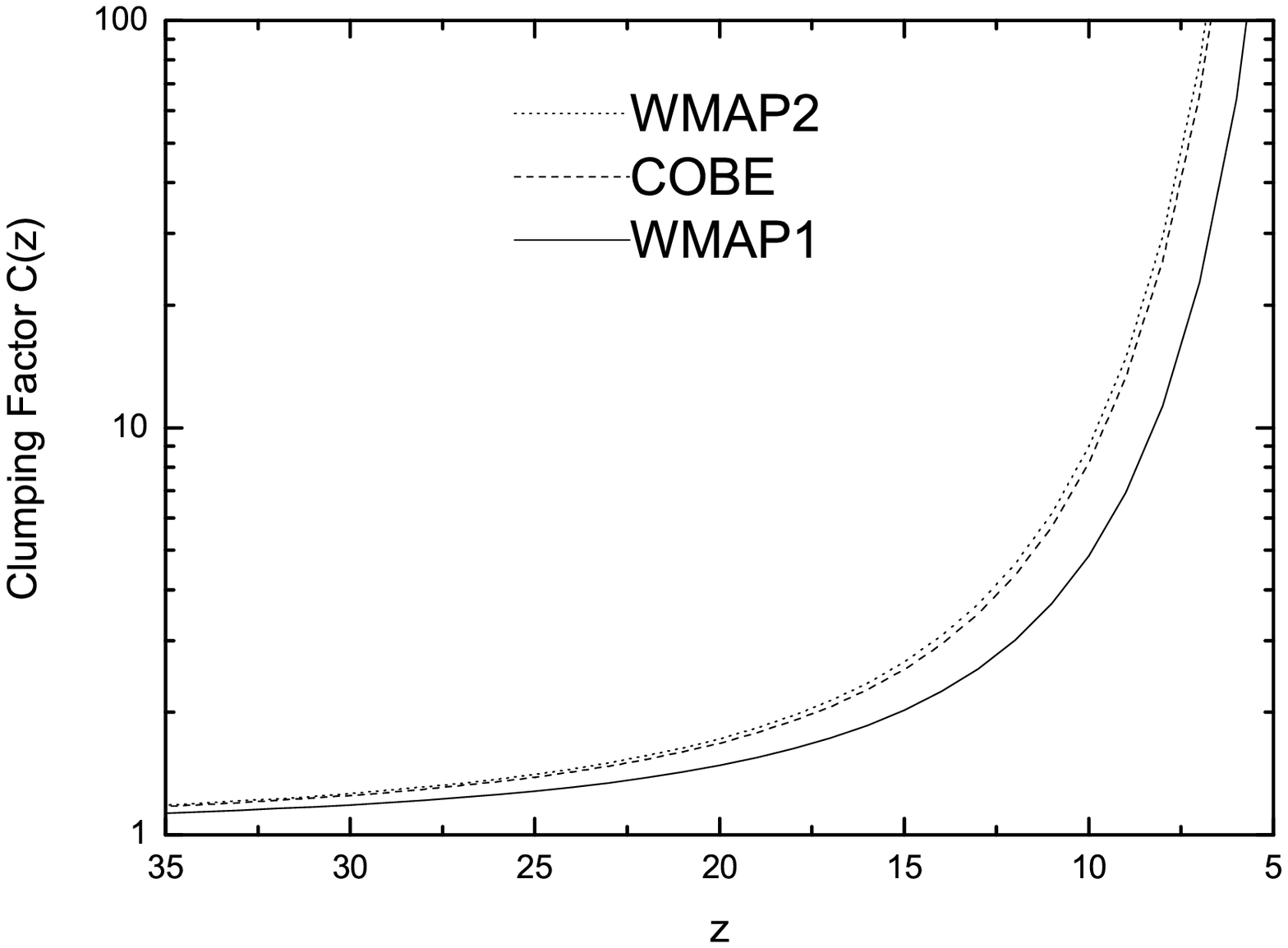} \figcaption {The clumping factor C(z) for the
three cosmological models given in Table 1.} \label{Fig3}

\plotone{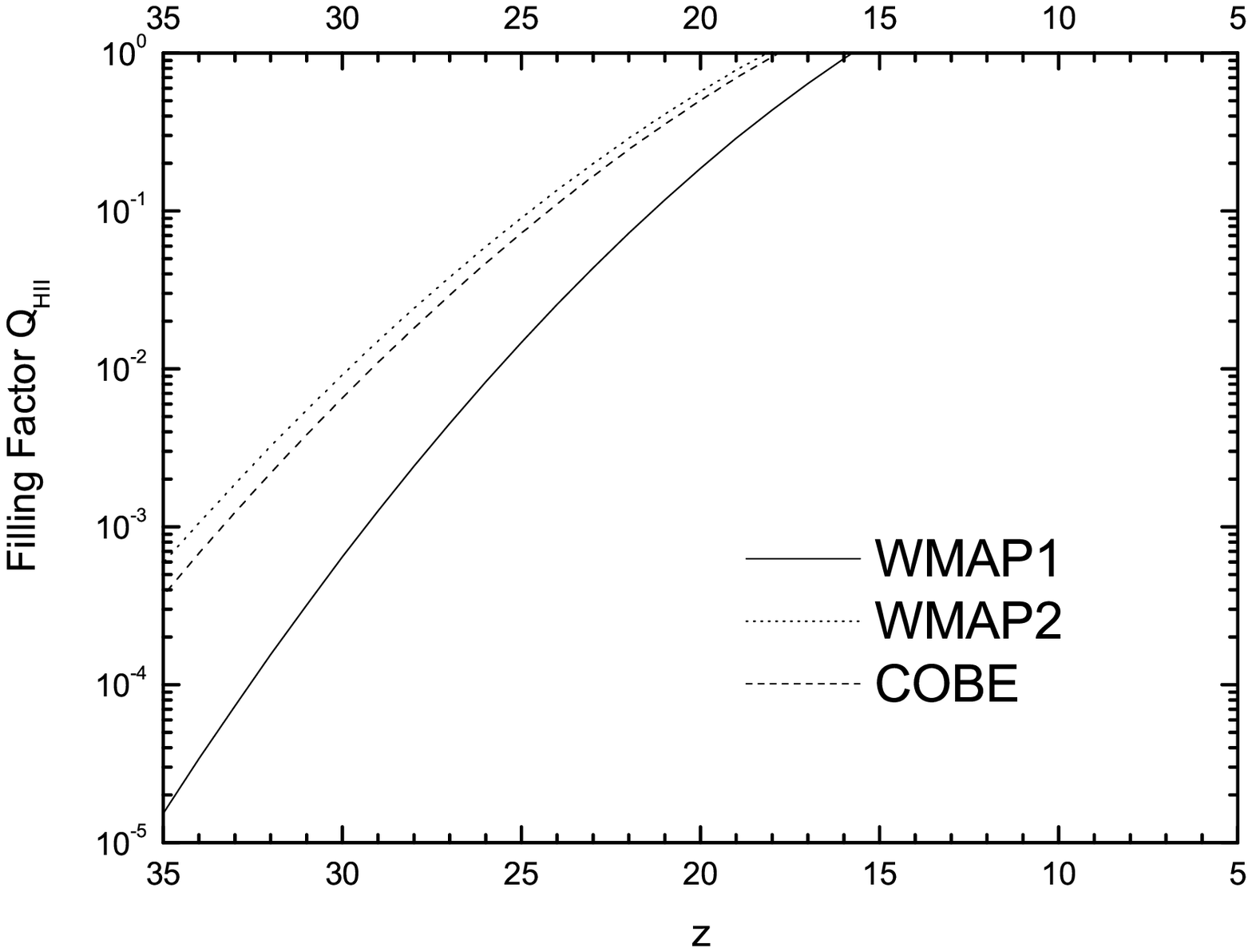} \figcaption {The filling factor $Q_{HII}$ vs.
redshift of the three cosmological models, WMAP1, WMAP2 and COBE.
$\rho/\bar{\rho}$ and $N_{ion}$ are taken to be 6 and 40,
respectively.
 } \label{Fig4}

\plotone{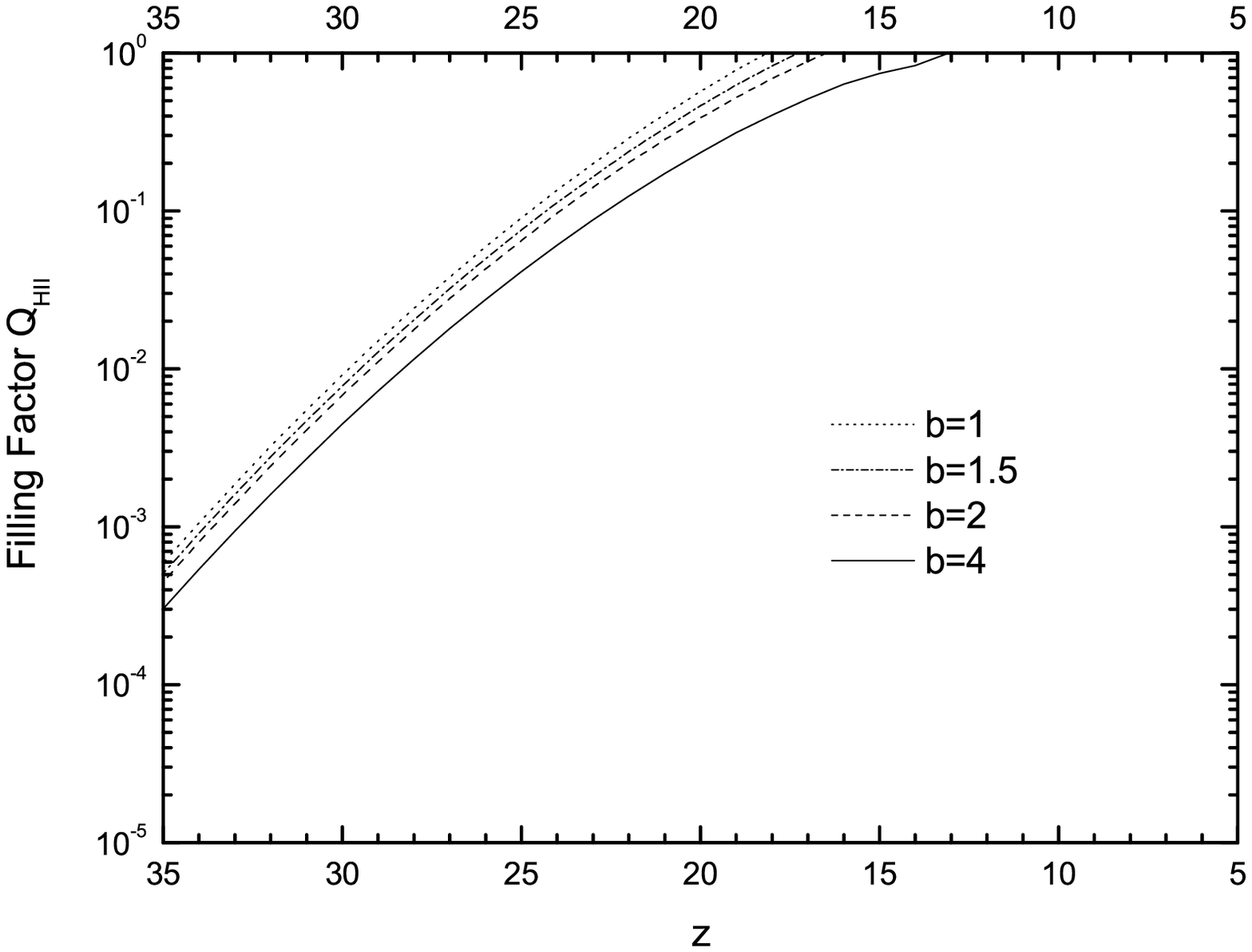} \figcaption {The filling factor $Q_{HII}$ vs.
redshift of the model WMAP2, $\rho/\bar{\rho}=6$ and $N_{ion}=40$.
The bias parameter $b$ of the clumping factor $C_{HII}=bC$ is
taken to be 1.5, 2 and 4.
 } \label{Fig5}

\plotone{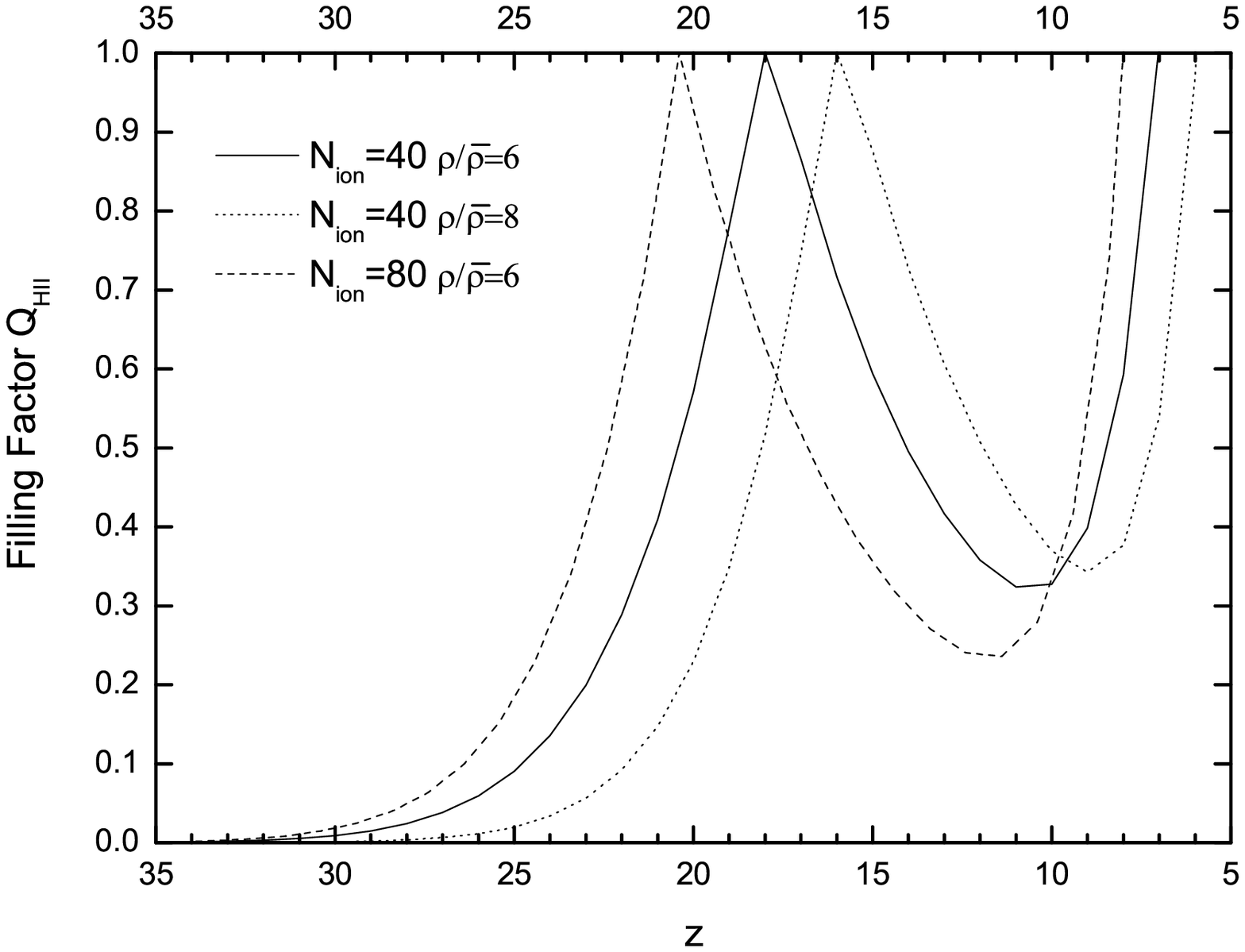} \figcaption {The redshift evolution of the volume
filling factor $Q_{HII}$(Z) from $z> z_r$ to $z_{gp}$ for model
WMAP2. $\rho/\bar{\rho}$ and $N_{ion}$ are taken to be (8,40)
(dot), (6,40) (solid), and (6,80) (dash). The reionization
redshifts $z_r$ is given by $Q_{HII}\approx1$ at $z>10$. The
redshift of the Gunn-Peterson transparency $z_{gp}$ is given by
$Q_{HII}\approx1$ at $z<10$. } \label{Fig6}

\end{document}